\documentclass[twocolumn,superscriptaddress,showpacs,preprintnumbers,amsmath,amssymb]{revtex4}

\usepackage[dvips]{graphicx}
\usepackage{dcolumn}
\usepackage{bm}
\newcommand{\msr}{$\mu$SR}
\newcommand{\lfaof}{LaFeAsO$_{1-x}$F$_x$}
\newcommand{\cfcaf}{CaFe$_{1-x}$Co$_x$AsF}
\newcommand{\bkfa}{Ba$_{1-x}$K$_x$Fe$_2$As$_2$}

\begin{document}

\preprint{APS/123-QED}
\title{Insular superconductivity in Co-doped iron pnictide \cfcaf\ }
\author{S. Takeshita}
\affiliation{Institute of Materials Structure Science, High Energy Accelerator Research Organization, Tsukuba, Ibaraki 305-0801, Japan}
\author{R. Kadono}
\affiliation{Institute of Materials Structure Science, High Energy Accelerator Research Organization, Tsukuba, Ibaraki 305-0801, Japan}
\affiliation{Department of Materials Structure Science, The Graduate University for Advanced Studies, Tsukuba, Ibaraki 305-0801, Japan}
\author{M. Hiraishi} 
\affiliation{Department of Materials Structure Science, The Graduate University for Advanced Studies, Tsukuba, Ibaraki 305-0801, Japan}
\author{M.~Miyazaki}
\affiliation{Department of Materials Structure Science, The Graduate University for Advanced Studies, Tsukuba, Ibaraki 305-0801, Japan}
\author{A. Koda}
\affiliation{Institute of Materials Structure Science, High Energy Accelerator Research Organization, Tsukuba, Ibaraki 305-0801, Japan}
\affiliation{Department of Materials Structure Science, The Graduate University for Advanced Studies, Tsukuba, Ibaraki 305-0801, Japan}
\author{S. Matsuishi}
\affiliation{Frontier Research Center, Tokyo Institute of Technology, 
Yokohama, Kanagawa 226-8503, Japan}
\author{H. Hosono}
\affiliation{Frontier Research Center, Tokyo Institute of Technology,
Yokohama, Kanagawa 226-8503, Japan}
\affiliation{Materials and Structures Laboratory, Tokyo Institute of Technology,
Yokohama, Kanagawa 226-8503, Japan}

%

\begin{abstract}

The presence of macroscopic phase separation between the superconducting and 
magnetic phases in \cfcaf\ is demonstrated by muon spin rotation (\msr) measurements conducted across their phase boundaries ($x=0.05$--0.15).  The magnetic phase tends to retain the high transition temperature ($T_{\rm m}>T_{\rm c}$), while Co-doping induces strong randomness.  The volumetric fraction of superconducting phase is nearly proportional to the Co content $x$ with constant superfluid density.  These observations suggest the formation of superconducting ``islands" (or domains) associated with Co ions in the Fe$_2$As$_2$ layers, indicating a very short coherence length.
\end{abstract}

\maketitle
The oxypnictide superconductor \lfaof\ (LFAO-F) with a critical temperature ($T_{\rm c}$)
of 26~K \cite{R_Kamihara} was recently discovered.  The successful revelation about the considerable increase in $T_{\rm c}$ on substitution of La with other rare-earth elements
(Ce, Pr, Nd, Sm, etc. leading to a maximum $T_{\rm c}$ of 55~K \cite{HCChen:08,ZARen:08,ZARen:08-2,GFChen:08}) and application of pressure for LFAO-F ($\sim$43~K \cite{R_Pressure}) has triggered wide interest in mechanisms that yield a relatively high $T_{\rm c}$ for this new class of compounds.   These compounds have a layered structure like that  exhibited by high-$T_{\rm c}$ cuprates, where the conducting Fe$_2$As$_2$ layers are isolated from the charge reservoir layers so that the doped carriers ({\sl electrons} in this case) introduced by the substitution of O$^{2-}$ with F$^{-}$ in the La$_2$O$_2$ layers can move within the layers of strongly bonded Fe and As atoms.  LFAO-F exhibit another qualitative similarity to cuprates: they exhibit superconductivity upon carrier doping of pristine compounds that exhibit magnetism \cite{Cruz:08,Nakai:08,Klauss:08,Carlo:08,McGuire:08,Kitao:08,Zhao:08,Aczel:08}.  The results of the recent muon spin rotation/relaxation ($\mu$SR) experiment conducted on a variety of iron-pnictide superconductors showed that the superfluid density $n_s$ falls on the empirical line on the $n_s$ vs $T_{\rm c}$ diagram of the {\sl underdoped} cuprates \cite{Carlo:08,Luetkens:08}; on the basis of this result, the possibility of the oxypnictides and cuprates having the same mechanism of superconductivity is discussed. 

The iron pnictides exhibit another interesting similarity with cuprates: the variation of their $T_c$ with doping concentration ($x$) is ``bell-shaped'' in hole-doped compounds ($A_{1-x}$K$_x$Fe$_2$As$_2$, $A=$ Ba, Sr) \cite{Rotter:08,Sasmal:08}, while $T_{\rm c}$ does not vary significantly with $x$ in electron-doped compunds \cite{R_Kamihara,Luetkens:08-2}. Recent investigations on electron-doped ($n$-type) cuprates strongly suggest that such electron-hole ``asymmetry" is a consequence of  the difference in the fundamental properties of underlying electronic states between the hole-doped and electron-doped compounds; the $n$-type cuprates are more like normal Fermi liquids rather than doped Mott insulators \cite{KSatoh:08}.  This result is strongly supported by the fact that given that all the doped carriers participate in the Cooper pairs (as suggested experimentally), the insensitivity of $T_{\rm c}$ to the variation of $n_s$ ($\propto x$) cannot be reconciled with the above mentioned empirical linear relation ($n_s$ vs $T_c$); the result is also reasonably understood on the basis of  the conventional BCS theory, in which condensation energy is predicted to be independent of carrier concentration. More interestingly, the very recent revelation about superconductivity induced by substitution of Fe with Co in LFAO and other iron pnictides (the Co atoms serve as electron donors) brings out the sheer contrast between the oxypnictides and cuprates in terms of tolerance to distortions in the conducting layers \cite{Sefat:08,Sefat:08-2,Qi:08,Matsuishi:08}.

Considering close relationship between magnetism and superconductivity, it is suggested that a detailed investigation of how these two phases coexist (and compete) near the phase boundary will provide important clues for the elucidation of the pairing mechanism. Among the various techniques available for this, $\mu$SR has an important advantage in that it can be applied to systems consisting of spatially inhomogeneous multiple phases to provide information on each phase, depending on their fractional yield. We conducted $\mu$SR measurements on \cfcaf\ (CFCAF, a variation of LFAO with trivalent cation and oxygen replaced with divalent alkali metal and fluorine, respectively, with the carrier doping achieved by substituting Co for Fe) and found a unique character of the Fe$_2$As$_2$ layers; it was found that the superconducting state exists over the vicinity of Co atoms, as inferred from the observation that the superconducting volume fraction is nearly proportional to the Co concentration, while $n_s$ remains unchanged. The rest of the CFCAF specimen exhibits magnetism (strongly modulated spin density wave),  indicating that superconductivity coexists with magnetism in the form of a phase separation. 

\begin{figure}[tp]
\begin{center}
\includegraphics[width=0.5\textwidth,clip]{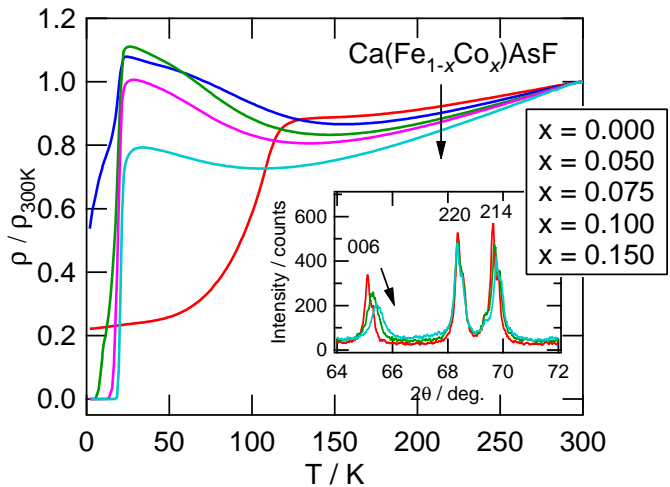}
\caption{(Color online)
Temperature dependence of electrical resistivity in \cfcaf\ with $x=0$, 0.05, 0.075, 0.10, and 0.15 normalized by the value at 300~K. (Inset) X-ray diffraction pattern around 006 peak for $x=0$, 0.075, and 0.15 (order of arrow direction).
}
\label{cf-rho}
\end{center}
\end{figure}

Unlike other oxypnictides with rare-earth metals, CFCAF has a major advantage because it is possible to identify the origin of magnetism, if at all detected by \msr, in the Fe$_2$As$_2$ layers without ambiguity. With the target concentration set around the phase boundary, polycrystalline samples with $x=0$, 0.05, 0.075, 0.10, and 0.15 have been synthesized by solid state reaction. The detailed procedure of sample preparation is the same as that described in an earlier report \cite{Matsuishi:08}, except that the sintering was carried out at 900 $^\circ$C for 20 h.  The samples were confirmed to be mostly of single phase by the X-ray diffraction method; CaF$_2$ (2.6, 3.3, and 6.2 wt\% for $x=0$, 0.075, and 0.15, respectively) and FeAs (3.3 wt\% for only $x=0.15$) were identified as the major impurities. It is known that the muons in fluorite exhibit a spin precession under zero field, which is characteristic of the F-$\mu$-F complex ($\simeq0.5$ MHz) \cite{Brewer:86}; the absence of such a signal indicates that the contribution of muons in fluorite is negligible. As shown in Fig.~\ref{cf-rho}, the pristine compound exhibits an anomaly in resistivity around 120 K, while the others (except for $x=0.05$) exist in the superconducting state below $T_c\simeq18$--21 K (defined as the midpoint of the fall in the resistivity). The homogeneity of the samples is supported by two findings: (i) the gradient of normalized resistivity above $\sim$150 K increases monotonously with increasing Co content $x$ (Fig.~\ref{cf-rho}), and (ii) the lattice parameter decreases linearly with increasing $x$ (Fig.~\ref{cf-rho}, inset).  

The conventional $\mu$SR measurement was performed using the
LAMPF spectrometer installed on the sample with $x=0$, 0.075 and 0.15 at the M20 beamline at TRIUMF, Canada. Additional data were obtained for the samples with $x=0.05$ and 0.10 using a new apparatus installed at the D1 beamline at the J-PARC MUSE Facility, Japan. In the measurements under zero field (ZF), the residual magnetic field at the sample position was reduced to below $10^{-6}$~T with the initial muon spin direction parallel to the muon beam direction [$\vec{P}_\mu(0)\parallel \hat{z}$].  For the longitudinal field (LF) measurements, a magnetic field was applied parallel to $\vec{P}_\mu(0)$. The time-dependent muon polarization [$G_z(t)=\hat{z}\cdot \vec{P}_\mu(t)$] was monitored by measuring the decay-positron asymmetry along the $\hat{z}$-axis.  The transverse field (TF) condition was achieved by rotating the initial muon polarization such that $\vec{P}_\mu(0)\parallel\hat{x}$, where the asymmetry was monitored along the $\hat{x}$-axis to obtain $G_x(t)=\hat{x}\cdot \vec{P}_\mu(t)$.  All measurements under magnetic field were performed by cooling the sample to the target temperature, after the field was equilibrated.

\begin{figure}[tb]
\begin{center}
\includegraphics[width=0.5\textwidth]{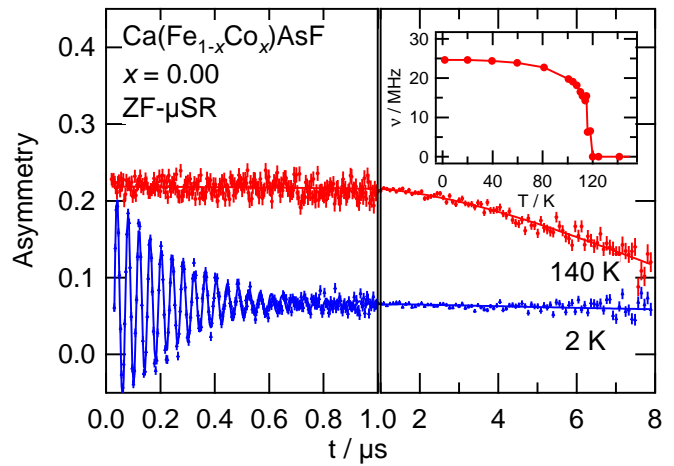}
\caption{(Color online)
ZF-\msr\ time spectra of CaFeAsF (undoped) at 140 K and 2 K. 
(inset) Frequency of spontaneous oscillation against temperature.
}
\label{cf-zf0}
\end{center}
\end{figure}

From the ZF-\msr\ measurement of the pristine compound ($x=0$), it was inferred that the anomaly around 120 K corresponds to the occurrence of magnetic phase transition. As shown in Fig.~\ref{cf-zf0}, the \msr\ spectra below $T_{\rm m}\simeq120$ K exhibit a spontaneous oscillation with a well-defined frequency, which approaches $\nu\simeq25$ MHz with decreasing temperature.  This indicates that the implanted muons sense a unique internal magnetic field of $B_{\rm m}=2\pi \nu/\gamma_\mu\simeq0.18$ T.  The magnitude of $B_{\rm m}$ is in good agreement with those reported in earlier \msr\ measurements of $R$FeAsO \cite{Klauss:08,Carlo:08,Aczel:08}, where it was suggested that a commensurate spin density wave (SDW) with a reduced moment of $\sim$0.25~$\mu_B$ exists at the iron sites \cite{Klauss:08}. In addition, from the LF-\msr\ spectra, it was inferred that the internal field is static within the time scale of \msr\ ($<10^{-5}$ s). 

\begin{figure}[tp]
\begin{center}
\includegraphics[width=0.45\textwidth,clip]{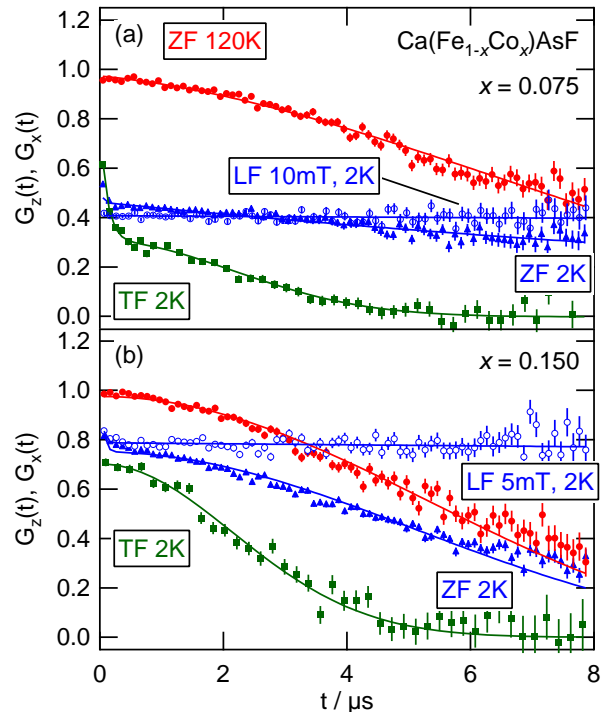}
\caption{(Color online) $\mu$SR time spectra observed in \cfcaf\ [$x=0.075$ (a) and 0.15 (b)] at 2 K under a longitudinal field (LF, open circles), a zero field (ZF, triangles), and a transverse field (TF, squares), and that under ZF above $T_m$ (triangles). The spectrum under TF is plotted on a rotating reference
frame to extract the envelop function.
}
\label{cf-tsp}
\end{center}
\end{figure}

\begin{figure}[tp]
\begin{center}
\includegraphics[width=0.45\textwidth,clip]{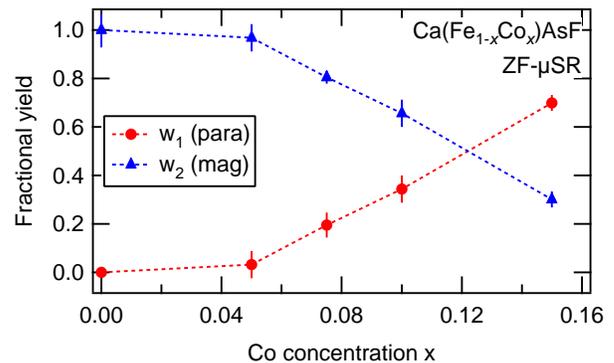}
\caption{(Color online) Relative yield of volumetric fraction of superconducting ($w_1$) and magnetic ($w_2$)   components in \cfcaf\ as a function of Co concentration ($x$).
}
\label{cf-frac}
\end{center}
\end{figure}

It has been reported that Co doping effectively suppresses the anomaly in the resistivity at $T_{\rm m}$; the anomaly virtually disappears at $x\simeq 0.1$, where the superconductivity seems to be close to its optimum, as suggested from the maximal $T_c\simeq22$ K \cite{Matsuishi:08}.  However, the ZF-\msr\ measurements of the samples with $x>0.05$ indicate that the superconductivity does not uniformly develop over the specimen. As shown in Fig.~\ref{cf-tsp}, the time spectra exhibit a character similar to that observed in the case of LFAO-F ($x=0.06$) \cite{Takeshita:08}; that is, they consist of two components, one showing rapid depolarization and the other showing slow Gaussian damping, with the relative yield of the latter increasing progressively with $x$.  A closer look at the initial time range of the spectra of the sample with $x=0.075$ shows that the rapid depolarization corresponds to a strongly damped oscillation with a frequency roughly equal to $\nu$.  This oscillation, together with the consistent onset temperature for magnetism ($T_{\rm m}\simeq120$ K), confirms that the signal arises from the SDW phase with a strong modulation caused by Co doping.  The very recent neutron diffraction experiment showed a similar result for $x=0.06$; however, only a volume-averaged signal was observed \cite{Xiao:08}. As observed in the TF-\msr\ measurements (see below), the rest of the samples exhibit superconductivity below $T_c$.  Considering these results, the ZF-\msr\ spectra are analyzed by the $\chi$-square minimization fit using a model function 
\begin{eqnarray}
G_z(t) &=&\left[w_1+w_2G_{\rm m}(t)\right]G_{\rm KT}(\delta_{\rm N}:t)\\ \label{E_ZF2}
G_{\rm m}(t)&=& \frac{1}{3} + \frac{2}{3}e^{-\Lambda_{\rm m} t}\cos(2\pi \nu t+\phi),\label{E_mag}
\end{eqnarray}
where  $G_{\rm KT}(\delta_{\rm N}:t)$ is the Kubo-Toyabe relaxation
function that describes the Gaussian damping due to random local fields
generated by nuclear moments (with $\delta_{\rm N}$ being the depolaization rate)
\cite{R_Hayano}, $w_1$ is the fractional yield for the nonmagnetic phase, $w_2$ is that for the SDW phase ($\sum w_i=1$), $\Lambda_{\rm m}$ is the depolarization rate for the spontaneous oscillation, and $\phi$ is the initial phase of rotation ($\simeq0$). The first term in Eq.~(\ref{E_mag}) represents the spatial average of $\cos\theta$, where $\theta$ is the angle between $\vec{P}_\mu(0)$  and $\vec{B}_{\rm m}$ at the muon site, which is equal to 1/3 in a polycrystalline specimen under zero external field. (This term would also be subject to depolarization under a fluctuating local field). The fractional yields of the respective components are shown in Fig.~\ref{cf-frac}. Depending on the magnitude of the internal field in the magnetically ordered phase of FeAs impurity, $w_2$  in the sample with $x=0.15$ might have a small uncertainty ($\simeq3.3$\%) below $T_{\rm N}\simeq77$ K.

\begin{figure}[tp]
\begin{center}
\includegraphics[width=0.45\textwidth,clip]{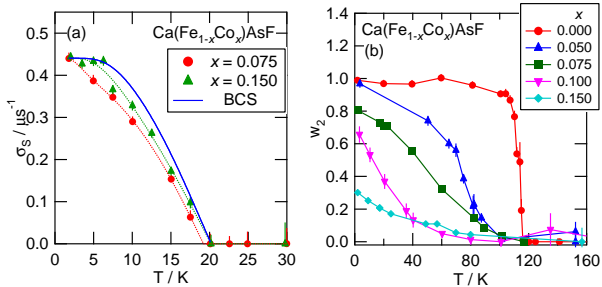}
\caption{(Color online)
Temperature dependence of (a) superfluid density ($\sigma_{\rm s}=\sqrt{2}\delta_{\rm s}$) and (b) fractional yield of magnetic phases for each sample.  The solid curve in (a) (labeled as BCS) represents the profile predicted using a weak-coupling BCS model with $s$-wave pairing and single gap. 
}
\label{G_TFmulti2}
\end{center}
\end{figure}

To analyze the temperature-dependent TF spectra, we used a model function
\begin{equation}
 G_x(t)= [w_1e^{-\delta^2_{\rm s}t^2}\cos(2\pi f_{\rm s} t+\phi)
 +w_2G_{\rm m}(t)]e^{-\frac{1}{2}\delta^2_{\rm N}t^2},\label{E_TFG}
\end{equation}
where $w_i$ and $\delta_{\rm N}$ are fixed as the values obtained by analyzing ZF- and LF-$\mu$SR spectra.  The first component in the above equation denotes the contribution of the flux line lattice formation in the superconducting phase, where $2\pi f_s\simeq\gamma_\mu H$ under an external field $H$, $\delta_{\rm s}$ corresponds to the linewidth $\sigma_{\rm s}=\sqrt{2}\delta_{\rm s}=\gamma_\mu\langle
(B({\bf r})-B_0)^2\rangle^{1/2}$, and $B_0\simeq H$ is the mean value of the local field $B({\bf r})$ \cite{R_Brandt}. The second term represents the relaxation in the magnetic phase.  The fit analysis using the above model indicates that all the spectra are perfectly reproduced when the fractional yield is fixed as the value determined from the ZF-$\mu$SR spectra.  This supports the assumption that the paramagnetic phase becomes superconducting below $T_{\rm c}$. The obtained values of $\sigma_{\rm s}$ are shown in Fig.~\ref{G_TFmulti2}(a).

The magnetic phase develops at temperatures much higher than the superconducting transition temperature ($T_{\rm m}>T_{\rm c}$), and in the case of $x\le0.075$, $T_{\rm m}$ almost remains unchanged  [see Fig.~\ref{G_TFmulti2}(b)].  Meanwhile, the oscillation observed in Fig.~\ref{cf-zf0} disappears in the spectra of all the Co-doped samples (Fig.~\ref{cf-tsp}, ZF, 2 K), indicating that the SDW state is strongly modulated. This observation, together with the absence of the \msr\ signal expected for ferromagnetic cobalt (4--5 MHz  \cite{Nishida:78}) in the spectra serves as evidence that the phase separation is not merely due to the aggregation of cobalt atoms during sample preparation.

As observed from its temperature dependence curves, shown in Fig. 5(a), $\sigma_{\rm s}$ ($\propto n_{\rm s}$) is almost independent of $x$.  Considering that the volume fraction of the superconducting phase ($w_1$) is nearly proportional to $x$, the insensitivity of $n_{\rm s}$ to $x$ indicates that the superfluid (and corresponding carrier density in the normal state) is confined to certain domains (``islands") centered around Co ions. A crude estimation showed that the domain size may be given as $d_s\sim(abc/2\cdot0.8/0.15)^{1/3}\simeq0.9$ nm in diameter (where $a$, $b$, and $c$ are the unit cell sizes).  In other words, the superfluid behaves as an incompressible fluid in CFCAF.  

The temperature dependence of $\sigma_{\rm s}$ shown in Fig.~\ref{G_TFmulti2}(a) is compared to the weak-coupling BCS model (s-wave, single gap).  The model fails to reproduce the present data for the cases of both $x=0.075$ and 0.15, as they exhibit a tendency of $\sigma_{\rm s}$ to vary with temperature over the region of $T/T_{\rm c}<0.4$. A very similar result is reported in the case of LFAO-F near the phase boundary ($x=0.06$) \cite{Takeshita:08}.  This suggests that provided the influence of flux pinning is negligible, the superconducting order parameter in CFCAF may not be explained by the simple weak-coupling BCS model with $s$-wave pairing and single-gap parameter.

The volumetric expansion of superconducting domains on electron doping to the Fe$_2$As$_2$ layers by substituting Fe with Co is a remarkable feature with no counterpart in high-$T_c$ cuprates.  Nonetheless, this feature, to some extent, is reminiscent of the  parallelism observed in cuprates on the substitution of Cu with Zn; it appears that superconductivity is suppressed over a certain domain around the Zn atoms like a ``Swiss cheese" \cite{Uemura:03}.  Although the effect discussed in the case of cuprates is completely opposite to that in the case of iron pnictides, the observed ``local" character of doping in CFCAF, which appears to originate from a short coherence length $\xi_0$ (that probably determines the domain size, so that $\xi_0\sim d_s/2\simeq0.45$ nm), may provide a hint for the microscopic understanding of superconductivity on the Fe$_2$As$_2$ layers, particularly in $n$-type doping.

Moreover, the superconducting character of $p$-type iron pnictides seems to be considerably different from that of the $n$-type ones, as suggested by the behavior of superfluid density of \bkfa\ observed by \msr\ \cite{Hiraishi:08}. The double-gap feature revealed by the angle-resolved photo-emission spectroscopy supports the view that  superconductivity occurs on complex Fermi surfaces consisting of many bands (at least five of them) which can give rise to certain intricacies \cite{Mazin:08,Kuroki:08}.  The peculiar feature of Co doping in CFCAF might also be understood in this context.

 We would like to thank the staff of TRIUMF and J-PARC MUSE for their technical support in the $\mu$SR experiment. This study was partially supported by the KEK-MSL Inter-University Program for Overseas Muon Facilities and by a Grant-in-Aid for Creative Scientific Research on Priority Areas from the Ministry of Education, Culture, Sports, Science and Technology, Japan.

\end{document}